\documentclass[conference]{IEEEtran}
\IEEEoverridecommandlockouts
\pdfoutput=1
\usepackage{cite}
\usepackage{amsmath,amssymb,amsfonts}
\usepackage{graphicx}
\usepackage{textcomp}
\usepackage{xcolor}
\usepackage{enumitem}
\usepackage{stfloats}

\def\BibTeX{{\rm B\kern-.05em{\sc i\kern-.025em b}\kern-.08em
    T\kern-.1667em\lower.7ex\hbox{E}\kern-.125emX}}

\usepackage{algorithm}
\usepackage[noend]{algpseudocode}

\usepackage[T1]{fontenc}
\usepackage[utf8]{inputenc}
\usepackage{soulutf8}
\usepackage{color}
\usepackage{svg}
\usepackage{verbatim}

\usepackage{array}
\usepackage{booktabs} 
\usepackage{multirow}

\usepackage{numprint}
\usepackage{nameref}
\usepackage{url}
\usepackage[hidelinks]{hyperref}
\usepackage{amsmath}
\usepackage{amsfonts}
\usepackage{amssymb}
\usepackage{xcolor}
\usepackage{float}
\usepackage{hyperref}
\usepackage{pifont}

\usepackage[T1]{fontenc}
\usepackage{colortbl}

\makeatletter
\providecommand{\tabularnewline}{\\}
\makeatother
\usepackage{babel}
\definecolor{Gray}{gray}{0.85}
\definecolor{LightGray}{gray}{0.95}
\definecolor{LightCyan}{rgb}{0.88,1,1}

\newcolumntype{L}[1]{>{\raggedright\arraybackslash}m{#1}}
\newcolumntype{C}[1]{>{\centering\arraybackslash}m{#1}}
\newcolumntype{R}[1]{>{\raggedleft\arraybackslash}m{#1}}

\usepackage{ragged2e}

\usepackage{longtable} 
\newcolumntype{L}[1]{>{\raggedright\let\newline\\\arraybackslash\hspace{0pt}}p{#1}}

\usepackage[strict]{changepage}
\usepackage{framed}
\definecolor{formalshade}{rgb}{0.95,0.95,1}

\newenvironment{formal}{%
  \MakeFramed{\advance\hsize-\width\FrameRestore}%
  \noindent\hspace{-4.55pt}
  \begin{adjustwidth}{}{7pt}%
}
{%
  \end{adjustwidth}\endMakeFramed%
}

\usepackage{color}
\definecolor{comments}{rgb}{0.13,0.55,0.13}
\definecolor{background}{rgb}{0.94, 0.97, 1.0}
\definecolor{strings}{rgb}{0.63,0.125,0.94}
\usepackage{listings}

\usepackage{lipsum}
\usepackage{tikz}
\usepackage{scalerel}
\usepackage[most]{tcolorbox}
\usepackage{setspace}
\usepackage{indentfirst}

\tcbset{
    enhanced,
    breakable,
    left=1mm,
    right=1mm,
    attach boxed title to top left={
        xshift=0.5cm,
        yshift= -3.5mm
    },
    top=3mm,
    bottom=0pt,
    colframe=blue!75!black,
    coltitle=white,
    colbacktitle=black!70!white,
    fonttitle=\footnotesize\bfseries,fontupper=\footnotesize\sffamily,
}

\makeatletter
\def\lst@makecaption{%
  \def\@captype{table}%
  \@makecaption
}

\usepackage{tikz}
\usepackage{enumitem}   

\newcommand*\circled[1]{\tikz[baseline=(char.base)]{
            \node[shape=circle,draw,inner sep=1pt,font=\sffamily\footnotesize] (char) {#1};}}

\begin{document}

\title{Towards Assessing Isolation Properties in Partitioning Hypervisors}

\author{\IEEEauthorblockN{Carmine Cesarano, Domenico Cotroneo, Luigi De Simone}
\IEEEauthorblockA{\textit{DIETI - Università degli Studi di Napoli Federico II, Via Claudio 21, 80125 Napoli, Italy}\\
\{carmine.cesarano\}@studenti.unina.it,\{cotroneo, luigi.desimone\}@unina.it}
}

\maketitle

\IEEEpubidadjcol

\begin{abstract}
Partitioning hypervisor solutions are becoming increasingly popular, to ensure stringent security and safety requirements related to isolation between co-hosted applications and to make more efficient use of available hardware resources. However, assessment and certification of isolation requirements remain a challenge and it is not trivial to understand what and how to test to validate these properties. Although the high-level requirements to be verified are mentioned in the different security- and safety-related standards, there is a lack of precise guidelines for the evaluator. This guidance should be comprehensive, generalizable to different products that implement partitioning, and tied specifically to lower-level requirements. The  goal of this work is to provide a systematic framework that addresses this need.
\end{abstract}


\section{Introduction}

In recent years, the development of technological solutions that exploit hardware more efficiently has become stronger. The current COVID-19-induced silicon shortage \cite{bloomberg_chip_shortage} has also contributed to this need. Generally, the industry brings the development of mixed-criticality systems (MCSs) to meet this requirement. With an MCS, several \textit{domains} at different levels of criticality are deployed into a common hardware platform, reducing hardware size and weight, cost, and power consumption. Virtualization technologies, along with the concept of \textit{partitioning}, are becoming a prominent way for the industry to consolidate multiple software systems on the same System-on-a-Chip in a flexible way \cite{blackberry, cinque2021virtualizing}. Partitioning hypervisors must ensure stringent isolation properties between the different coexistent domains. The isolation property allows individual domains to be isolated by making them independent so that they cannot interfere with each other. Isolation must be considered both in spatial and temporal terms, as well as in terms of fault containment. 

In addition, when partitioning solutions are used in certified systems, it is necessary to consider the above properties in light of security or safety certification requirements.
The main difference between safety and security is the origin of the risk. Safety considers hazards and, thus, accidental component failures or software errors, while security considers threats and focuses on intentional malicious attacks. Despite these differences, it is important to note that both attacks and incidents can cause damage to system assets in terms of people, property, environments, or services. Suffice to remember the famous Stuxnet \cite{chen2011lessons} and Flame \cite{mcelroy2012flame} attacks in safety-critical domains. Therefore, safety and security requirements should be clearly treated in critical contexts \cite{brygier2009safe, pfitzmann2004safety, bohner2015extending, li2011merging}. As for a traditional virtualization solution, performing safety/security assessment of a partitioning hypervisor in a manageable and a cost-effective way remains a big challenge. 

Existing work, both from industry and from academia, recommends the use of manual tests \cite{michaels2019assessing, cotroneo2021timing}, fault injection \cite{gunawi2011fate, joshi2011prefail, ju2013fault, cerveira2020effects}, or fuzzing testing \cite{fonseca2018multinyx, schumilo2020hyper, schumilo2021nyx, ge2021hyperfuzzer} to assess some components (e.g., device emulation, memory virtualization, CPU virtualization). As part of a certification process, pre-existing tools are insufficient. First, individually they do not cover all the isolation mechanisms of the partitioning solution under test. Further, they also provide no evidence on the certification requirements that may have been covered by the tests. Other studies include the use of formal methods to verify the properties of a partitioning solution \cite{baumann2011proving, cohen2009vcc, richards2010modeling, heitmeyer2006formal}. Despite formal verification has proven to be a good way to assess isolation requirements, it makes the process costly. In addition, the use of these techniques is mandatory only for high-assurance levels and not for intermediate ones. Moreover, there is currently no approach that is easily generalizable to different partitioning solutions and directly addresses certification requirements. 

The objective of this work is to move toward filling this gap. We propose a comprehensive framework to evaluate the isolation level of a partitioning hypervisor, considering together the security and safety certification requirements. Considering requirements from different standards can make the certification process more efficient and less costly, allowing the reuse of any artifacts produced. 

The contributions of this work are as follows:

\begin{itemize}
    
    \item Analysis of isolation properties highlighted in certification standards to propose a mapping between the functional mechanisms implementing isolation requirements in a virtualization solution and the certification functional requirements; An example mapping is provided considering the ISO 15408 standard and the requirements described in the Separation Kernel Protection Profile (SKPP);
    \vspace{-1cm}
    \bigbreak
    \item A comprehensive safety and security methodological framework to assess isolation properties of partitioning hypervisors for certification purposes.
    
\end{itemize}


\pagebreak
\section{Background}
\label{sec:backgroud}

\subsection{Partitioning concepts and virtualization solutions}
The concept of partitioning was born to isolate different processes running on common hardware, in terms of spatial, temporal, and fault isolation, making them appear independent. \textbf{Partition}, also known as domain, represents the logical isolated unit. It is an abstraction that separates a portion of a certain resource (such as CPU time or memory) from all other portions. From ISO 15408, DO-178B, and IEC 61508 standards:

\begin{itemize}
    \item \textbf{Temporal isolation} is the ability to isolate the impact of the usage of a resource (such as CPU or memory) of a certain partition on the decay of the performance of another partition. This avoids phenomena such as starvation or throughput reduction.
    \item \textbf{Spatial isolation} (also known as memory isolation) is the ability to isolate code and data of a partition from the others. This prevents data and code alteration or eavesdropping.
    \item \textbf{Fault isolation} prevents any failure in one partition from causing failures in another partition or impacting the overall system throughput. 
\end{itemize}

Over the years, all architectural solutions that have been designed to deal with partitioning are based on virtualization technologies. Virtualization provides a software layer that abstracts the hardware resources to run different and isolated application environments. Virtualization approaches can be divided into full-virtualization or paravirtualization. Full-virtualization, which can also be hardware-assisted, allows the complete abstraction of the hardware resources (e.g., CPU, memory, etc.) to the guests, emulating privileged instructions and I/O operations. On the other hand, paravirtualization includes modifying the guest Operating System (OS) to communicate directly with the underlying hypervisor, through so-called hypercalls, a mechanism similar to system calls. \textit{Cotroneo et al.} \cite{cinque2021virtualizing} grouped the virtualization approaches used or proposed for industrial mixed-criticality systems into four main categories: 

\begin{itemize}
    \item Solutions based on separation kernel and microkernel, specifically designed for industrial and embedded domains (e.g., Lynx OS-178 \cite{lynxsecure}, PikeOS \cite{pikeos}, VxWORKS \cite{windrivervxworks}, Jailhouse \cite{ramsauer2017look}, Bao \cite{martins2020bao});
    
    \item Solutions that extend general-purpose hypervisors in order to support real-time properties. These solutions foster the adoption of mainstream virtualization solutions in the industry (e.g., DornerWorks Xen-based hypervisor named ARLX \cite{vanderleest2013safe}).
 
    \item Solutions based on the isolation support provided by security CPU hardware extensions (e.g., the SGX-based solution proposed by De Simone et al.\cite{de2019isolating}, or the ARM TrustZone-assisted hypervisor LTZvisor \cite{pinto2017lightweight});
    
    \item Solutions based on lightweight virtualization, such as containers or unikernels, which try to achieve a compromise between isolation and the small footprint required in some industrial domains (e.g., includeOS \cite{includeOS}, rumprun \cite{rumprun}, mirageOS \cite{mirageOS}).
\end{itemize}

Of all the virtualization solutions mentioned above, the ones that best fit the concept of partitioning are those based on the separation kernel. In fact, these solutions, also known as \textbf{partitioning hypervisors}, are specifically designed to ensure isolation. A partitioning hypervisor is developed with the smallest footprint possible, providing the least number of services, ensuring partition management and static allocation of resources (e.g. CPU and memory) to partitions. It must also ensure that no communication channels are established among the partitions other than the explicitly defined ones. The partitioning hypervisor is the virtualization solution addressed by the framework proposed in this paper.

\subsection{Partitioning mechanisms}
\textit{H. Blasum} \cite{blasum2015partitioning} analyzed different architectures, including general-purpose OS (GPOS), real-time OS (RTOS), Multiple Independent Level of Security kernel (MILS) and identified several functional mechanisms that contribute to spatial and temporal isolation properties. Some of them are already available in a GPOS, others can be found in a safety-compliant RTOS. In a partitioning hypervisor, there are the same mechanisms, properly enforced. 

\subsubsection{Memory partitioning mechanisms}
\begin{itemize}
    \item \textbf{[M1] Access control to user-space memory}: MMU configurations to ensure the separation of user-space memory portions among partitions, restricting accesses. This ensures memory integrity (control of writes) and memory confidentiality (control of reads) between two or more different partitions.
    \item \textbf{[M2] Access control to kernel-space memory}: MMU configurations to ensure the separation of kernel-space memory portions containing, for example, management data (upper bounds for resource usage, actual usage, the state of threads running in the kernel, etc.).
    \item \textbf{[M3] Access control to hardware resources}: kernel configurations to completely separate hardware resources among the partitions or, if the resources are shared, to separate their use by time windows. This enforces confidentiality.
    \item \textbf{[M4] Static memory allocation}: memory management mechanisms to statically allocate fixed quotas of memory to each partition. This ensures that no partition can deplete the storage space (for example, by a fork bomb), enforcing availability. 
\end{itemize}

\subsubsection{Temporal partitioning mechanisms}
\begin{itemize}
    \item \textbf{[T1] CPU registers reuse}:  kernel mechanism for temporal partitioning of CPU register between applications. This makes residual information unavailable on context switches, enforcing confidentiality. 
    \item \textbf{[T2] Cyclical Scheduler}: kernel mechanisms to impose fixed scheduling that ensures each partition gets access to the CPU on a cyclic basis in its time slices. This enforces availability.
    \item \textbf{[T3] Worst-case execution time}: kernel mechanism to ensure an upper bound of the worst-case execution time for each critical execution path in the system. These bounds could also refer to the resource utilization time of a task. This enforces readiness and availability, preventing a task from stalling another task.
    \bigbreak
    \item \textbf{[T4] Temporal normalization}: while WCET sets an upper bound for the time of resource utilization, temporal normalization (TN) imposes that the utilization time is fixed. It is possible to implement TN by inserting small empty time windows after useful work to fix the slot duration. This prevents resource modulation to establish a timing covert channel, enforcing confidentiality.
\end{itemize}

This asset-based analysis has supported the definition of the proposed methodological framework. In fact, the identification of partitioning mechanisms is important to better understand which components should be tested and monitored in a partitioning hypevisor solution.

\section{Isolation Properties in Standards}
\label{sec:isolation}

\begin{table*}[!ht]
\renewcommand*{\arraystretch}{1.2}
\caption{Spatial, temporal, and fault isolation properties in different standards}
\label{tab:isolation_in_standards}
\resizebox{\textwidth}{!}{%
\sffamily 
\footnotesize 
\setstretch{0.90}
\begin{tabular}{@{\extracolsep{\fill}}|>{\raggedright}m{.1\textwidth}|>{\raggedright}m{.06\textwidth}|>{\raggedright}m{.2\textwidth}|>{\raggedright}m{.07\textwidth}|>{\raggedright}m{.2\textwidth}|>{\raggedright}m{.07\textwidth}|>{\raggedright}m{.2\textwidth}|>{\raggedright}m{.07\textwidth}|}

\hline 

\rowcolor{blue!8}
\multirow{2}{3cm}{} 
& \multirow{2}{2cm}{} 
& \multicolumn{2}{c|}{\cellcolor{blue!25}\textbf{Spatial Isolation}} 
& \multicolumn{2}{c|}{\cellcolor{blue!35}\textbf{Temporal Isolation}} 
& \multicolumn{2}{c|}{\cellcolor{blue!45}\textbf{Fault Isolation}
}\tabularnewline
\cline{3-8} 

\rowcolor{blue!8}
\centering\textbf{Standard} 
& \centering\textbf{Scope}
& \centering\textbf{Details} 
& \centering\textbf{Ref.} 
& \centering\textbf{Details} 
& \centering\textbf{Ref.} 
& \centering\textbf{Details} 
& \centering\textbf{Ref.}
\tabularnewline

\hline 
\hline 
\textbf{DO-178C (Avionics) \cite{do178C}} 
& \textbf{\emph{Safety}}  
& "A partitioned software component should not be allowed to contaminate another partitioned software component’s code, input/output (I/O), or data storage areas."  
& Section 2.4.1.a
& "A partitioned software component should be allowed to consume shared processor resources only during its scheduled period of execution."
& Section 2.4.1.b
& "Failures of hardware unique to a partitioned software component should not cause adverse effects on other partitioned software components."
& Section 2.4.1.c
\tabularnewline
\cline{1-8} 

\hline 
\hline 

\textbf{IEC 61508, Part 3 (Generic) \cite{iec61508}} 
& \textbf{\emph{Safety}}  
& "Spatial: the data used by one element shall not be changed by another element. In particular, it shall not be changed by a non-safety-related element. 
& F.2, F.4 (Annex F)
& "Temporal: one element shall not cause another element to function incorrectly by taking too high a share of the available processor execution time or by blocking the execution of the other element by locking a shared resource of some kind."
& F.2, F.5 (Annex F)
& Not explicitly mentioned, but implied by "independence of execution" and "non-interference" of isolated applications.
& F.3 (Annex F)
\tabularnewline
\cline{1-8} 
\hline 

\hline 
\hline 

\textbf{ISO 26262, Part 6 (Automotive) \cite{iso26262}} 
& \textbf{\emph{Safety}}  
& "With respect to memory, the effects of faults such as those listed below can be considered for software elements executed in each software partition: corruption of content; read or write access to memory allocated to another software element."
& D.2.3 (Annex D)
& "With respect to timing constraints, the effects of faults such as those listed below can be considered for the software elements executed in each software partition: blocking of execution; deadlocks; livelocks; incorrect allocation of execution time; incorrect synchronization between software elements."
& D.2.2 (Annex D)
& Not explicitly mentioned, but implied by "freedom from interference" of isolated applications.
& D2.1 (Annex D)
\tabularnewline
\cline{1-8} 
\hline 

\hline 
\hline 

\textbf{EN 50128 (Railway) \cite{en50128}} 
& \textbf{\emph{Safety}}  
& Not explicitly mentioned, but implied by "Response Timing and Memory Constraints" 
& D.45 (Annex D)
& Not explicitly mentioned, but implied by "Response Timing and Memory Constraints" 
& D.45 (Annex D)
& N/A
& N/A
\tabularnewline
\cline{1-8} 
\hline 

\hline 
\hline 

\textbf{ISO 15408, Part 2 (Generic) \cite{commoncriteria}} 
& \textbf{\emph{Security}}  
& Not explicitly mentioned but implied by access control information flow.
& FDP\_ACC, FDP\_ACF
& Not explicitly mentioned, but implied by resource management.
& FRU\_RSA
& N/A
& N/A
\tabularnewline
\cline{1-8} 
\hline 

\end{tabular}
}
\end{table*}

As the concept of mixed-criticality is increasingly adopted in industry, spatial, temporal, and fault isolation must be considered in the certification process. Several international standards recommend verification activities to certify the isolation level provided by the system under test (SUT). For instance, fault injection testing, robustness testing, or performance testing are recommended. These activities are used during the certification process by the evaluator, but also by the developer to demonstrate the prerequisites for certification. \tableautorefname{}~\ref{tab:isolation_in_standards} shows how five different standards \cite{do178C, commoncriteria, iec61508, iso26262, en50128} treat or mention isolation properties. For each of the standards, the "\textit{Scope}" column gives indications about the type of standard (Safety or Security). The columns "\textit{Spatial isolation}", "\textit{Temporal Isolation}" and "\textit{Fault Isolation}" give details about the isolation properties and their references in the standard. In addition, some standards also mention specific techniques to ensure these properties. For example, DO-178B \cite{do178C}, IEC 61508 \cite{iec61508}, and ARINC-653 \cite{arinc653}, recommend temporal predictability using fixed cyclical scheduling, time-triggered scheduling, fixed priority-based scheduling, monitoring of CPU execution time, or WCET analysis for temporal isolation purposes. Instead, parity bits, error-correcting code, cyclic redundancy check, redundant storage, and restricted access to memory through MMU or IOMMU are common techniques recommended for achieving spatial isolation. It should be noted that temporal partitioning for safety requires that the hardware utilization of a partition does not affect the availability of resources of another partition. Temporal partitioning for security requires that a partition cannot detect when another partition is using or not the resource. On the other hand, regarding memory partitioning, safety standards may not consider confidentiality. For example, DO-178C states: 

\begin{formal}
\textbf{Section 2.4.1}: \textit{"A partitioned software component should not be allowed to contaminate another component's code, I/O, or data storage areas."} \cite{do178C}
\bigbreak
\bigbreak
\noindent \textbf{Section 2.3.1}:  \textit{"Memory partitioning is ensured by prohibiting memory accesses (at a minimum, write access) outside a partition's defined memory areas."}  \cite{do178C}
\end{formal}

The "\textit{not to contaminate}" requirement does not imply any measures to prevent eavesdropping. However, from a product perspective, if an access control mechanism has been implemented, this will control access not only for "writes", but also for "reads". One of the key factors in designing the proposed framework is the identification of the software components that provide the isolation features. The latter will be monitored and validated during the certification process. One more important factor is the identification of testing techniques to be used for validation. The factors described above must be mapped to the requirements addressed by the specific standard, in order to use the proposed framework in a certification context. So there is a need to provide the following mappings.

\begin{enumerate}[label=\protect\circled{M\textsubscript{\Alph*}}]
    \item \label{mapA}  \textit{Mapping between the partitioning mechanisms and the functional requirements specified in the standard.}
    \item \label{mapB}  \textit{Mapping between existing testing techniques and the assurance requirements specified in the standard.}
\end{enumerate}

References to the partitioning mechanisms identified \cite{blasum2015partitioning} are few or absent in a \textit{process-oriented standard} (often those related to safety). Besides, it is only possible to extrapolate higher-level functional requirements (see Table ~\ref{tab:isolation_in_standards}). In this case, the mapping \ref{mapA} cannot be provided only from the standard. Typically, the lower-level requirements of the specific product under test are provided by the developer, who can help define the above mapping. On the contrary, the mapping \ref{mapB} can be evaluated even with product unawareness. On the other hand, in a \textit{product-oriented standard} (such as the security-oriented ISO 15408 standard) one can infer both mappings. In fact, in this case, the standard provides both functional and assurance requirements for the precise class of product to be tested. 

\begin{figure}[!b] 
    \centering
    \includegraphics[width=1\columnwidth]{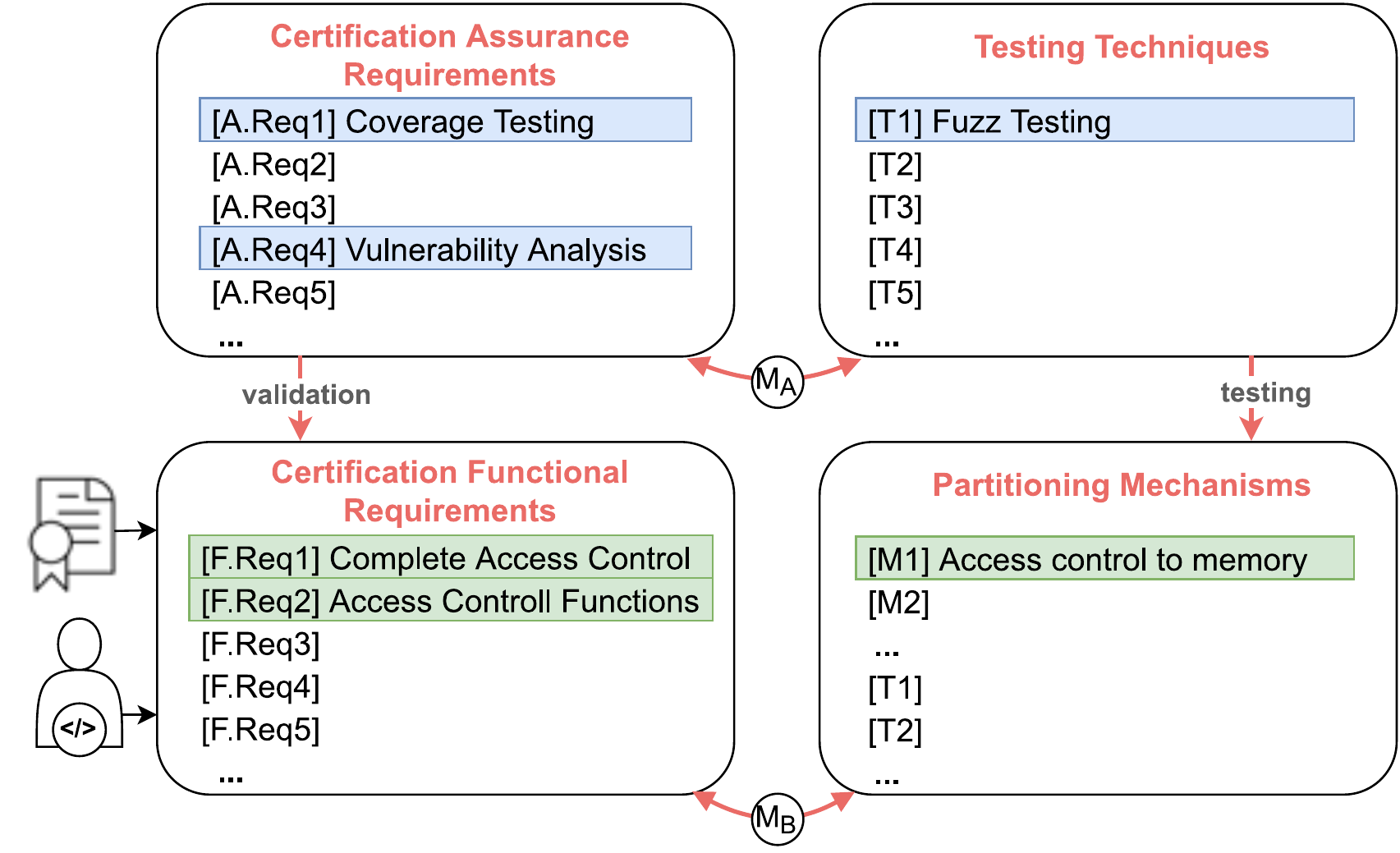}
    \caption{ISO 15408: example of the mappings \ref{mapA} and \ref{mapB}}
    \label{fig:mappings}
\end{figure}

\begin{table}[!b]
\renewcommand*{\arraystretch}{2}
\caption{ISO 15408: partitioning mechanisms and sfrs mapping}
\label{tab:mappingSFR}
\resizebox{.5\textwidth}{!}{%
\sffamily 
\footnotesize 
\setstretch{0.90}
\begin{tabular}{@{\extracolsep{\fill}}|>{\raggedright}m{.1\textwidth}|>{\raggedright}m{.3\textwidth}|>{\raggedright}m{.1\textwidth}|}
\hline 

\rowcolor{blue!8}
\centering\textbf{Ref.} & 
\centering\textbf{Security Functional Requirement class} & 
\centering\textbf{Separation Mechanism}
\tabularnewline

\hline 
\hline 
\textbf{\emph{[FDP\_IFC.2.1]}}
& \textbf{Information Flow Control Policy}  \\
The TSF\footnote{TSF: Target of Evaluation Security Functions} shall enforce the Partitioned Information Flow
SFP \footnote{SFP: Security Function Policy} on all partitions, all subjects, all exported resources
for all possible operations that cause information to flow between
subjects and exported resources.
& \textbf{\emph{[M1][M2][M3]}} 
\tabularnewline

\hline 
\hline 
\textbf{\emph{[FDP\_IFF.1.1]}}
& \textbf{Information Flow Control Functions} \\ 
The TSF shall enforce the Partitioned Information Flow SFP as a Partition Abstraction based on the flow(s) caused by an operation,
and the following types of partition, subject, and exported resource
security attributes associated with the operation: 
\begin{itemize}
    \item The identity of the subject involved in the flow of information;
    \item The identity of the partition to which the subject is assigned;
    \item The identity of the exported resource involved in the flow of information;
    \item The identity of the partition to which the exported resource is assigned. 
\end{itemize}
& \textbf{\emph{[M3]}}
\tabularnewline

\hline 
\hline 
\textbf{\emph{[FMT\_MOF]}}
& \textbf{Management of Functions} \\
The TSF shall restrict the ability to invoke a configuration change of the TOE, invoke a restart of the TOE, invoke a halt of the TOE, invoke a transition of the TOE to maintenance mode to authorized subjects
& \textbf{\emph{[M2]}}
\tabularnewline

\hline 
\hline 
\textbf{\emph{[FMT\_IFF.3.1]}}
& \textbf{Limited Illicit Information Flows} \\
The TSF shall enforce the Partitioned Information Flow SFP to limit
the capacity of covert timing channels and covert storage channels
between partitions to [assignment: metric establishing maximum
covert channel capacity]. 
&  \textbf{\emph{[M4][T4]}}
\tabularnewline

\hline 
\hline 

\textbf{\emph{[FDP\_RIP.2.1]}}
& \textbf{Full Residual Information Protection} \\
The TSF shall ensure that any previous information content of a resource is made unavailable upon the \textit{[selection: allocation
of the resource, deallocation of the resource]}.
& \textbf{\emph{[T1]}}
\tabularnewline

\hline 
\hline 
\textbf{\emph{[FTP\_SEP]}}
& \textbf{Domain Separation } \\
The unisolated portion of the TSF shall use hardware
mechanisms to maintain a security domain for its execution that
protects the code and data of the unisolated portion of the TSF from
interference and tampering by untrusted subjects.
&  \textbf{\emph{[M3]}}
\tabularnewline

\hline 
\hline 
\textbf{\emph{[FRU\_RSA]}}
& \textbf{Minimum and Maximum Quotas} \\
The TSF shall enforce minimum/maximum quotas of the following
resources for each partition, as defined by the configuration data:
\begin{itemize}
    \item System memory
    \item Processing time
\end{itemize}
& \textbf{\emph{[M4][T3][T4]}}
\tabularnewline

\hline 
\hline 
\textbf{\emph{[FRU\_PRU]}}
& \textbf{TSF Predictable Resource Utilization} \\
The TSF shall exhibit predictable and bounded execution behavior with respect to its usage of processor time and memory resources. 
&\textbf{\emph{[M4][T3][T4]}}

\tabularnewline

\hline 

\end{tabular}
}
\end{table}

Among certification standards, \textit{ISO 15408} (also known as Common Criteria) \cite{commoncriteria} is a good example for providing the above mappings, being a product-oriented standard. Moreover, over the years, the certification requirements related to isolation properties have already been formalized, for example, for the Separation Kernel architecture. ISO 15408 provides a standardized security framework that allows users to define unambiguously and comprehensively threats, security objectives, and assumptions related to an IT system, by writing a Protection Profile (PP) draft. The two basic elements of a PP are \textit{Security Functional Requirements} (SFRs), which are the set of security-related requirements to be verified during the product certification process, and \textit{Security Assurance Requirements} (SARs), which instead detail the recommended or mandated techniques for verifying SRFs. Figure ~\ref{fig:mappings} shows an example of the two mappings with reference to the ISO 15408 standard, making explicit the importance of these correspondences. From a product perspective, a given testing technique (for example, fuzzing) is applied to a functional mechanism (for example, to the access control of memory) to demonstrate its isolation properties. From a certification perspective, testing provides evidence about the fulfillment of a set of assurance and functional certification requirements. 
\tableautorefname{}~\ref{tab:mappingSFR}, on the other hand, shows mapping \ref{mapA} in detail, referring to the SFRs extracted from the draft of the "\textit{Protection Profile for Separation Kernels in Environments Requiring High Robustness}" \cite{skpp}. The Target of Evaluation (TOE) addressed by this PP is a Separation Kernel, an architecture that can easily be representative of partitioning hypervisors. With reference to the same PP, \tableautorefname{}~\ref{tab:mappingSAR} reports some of the most interesting SARs for \verb|ATE| and \verb|AVA| classes, showing the mapping \ref{mapB}. The \verb|ATE| class regards tests which encompass coverage, depth, independent testing, and functional tests. The \verb|AVA| class includes tests that try to find exploitable vulnerabilities introduced in development.

\section{Partitioning Hypervisor Assessment}
The asset-based analysis of partitioning hypervisors presented in Section \ref{sec:backgroud}, along with the analysis of the certification requirements related to isolation in the Section \ref{sec:isolation}, led to the definition of a methodological framework for the isolation assessment of a partitioning hypervisor solution. If the isolation properties are appropriately tested, the solution can be certified in terms of security and safety. Therefore, the proposed framework does not vertically address a specific component, but aims to benchmark the overall degree of isolation of a partitioning solution in order to answer the question: \textit{How much isolation does the target partitioning solution provide?}

The framework should support the following.

\begin{itemize}

    \item \textbf{Support certification standards:} fine-tuned to meet different security- and safety-related requirements and different declinations of isolation. 
    
    \item \textbf{Support for different partitioning solution:}  different products can implement, for example, different hypervisor interfaces.
    
    \item \textbf{Support different types of testing:} different target components may require different types of testing.
    
    \item \textbf{Limited impact on the system under test: } framework components must be modular and must not intrusively impact the hypervisor.
\end{itemize}

\begin{figure*}[!t] 
    \centering
    \includegraphics[width=1\textwidth]{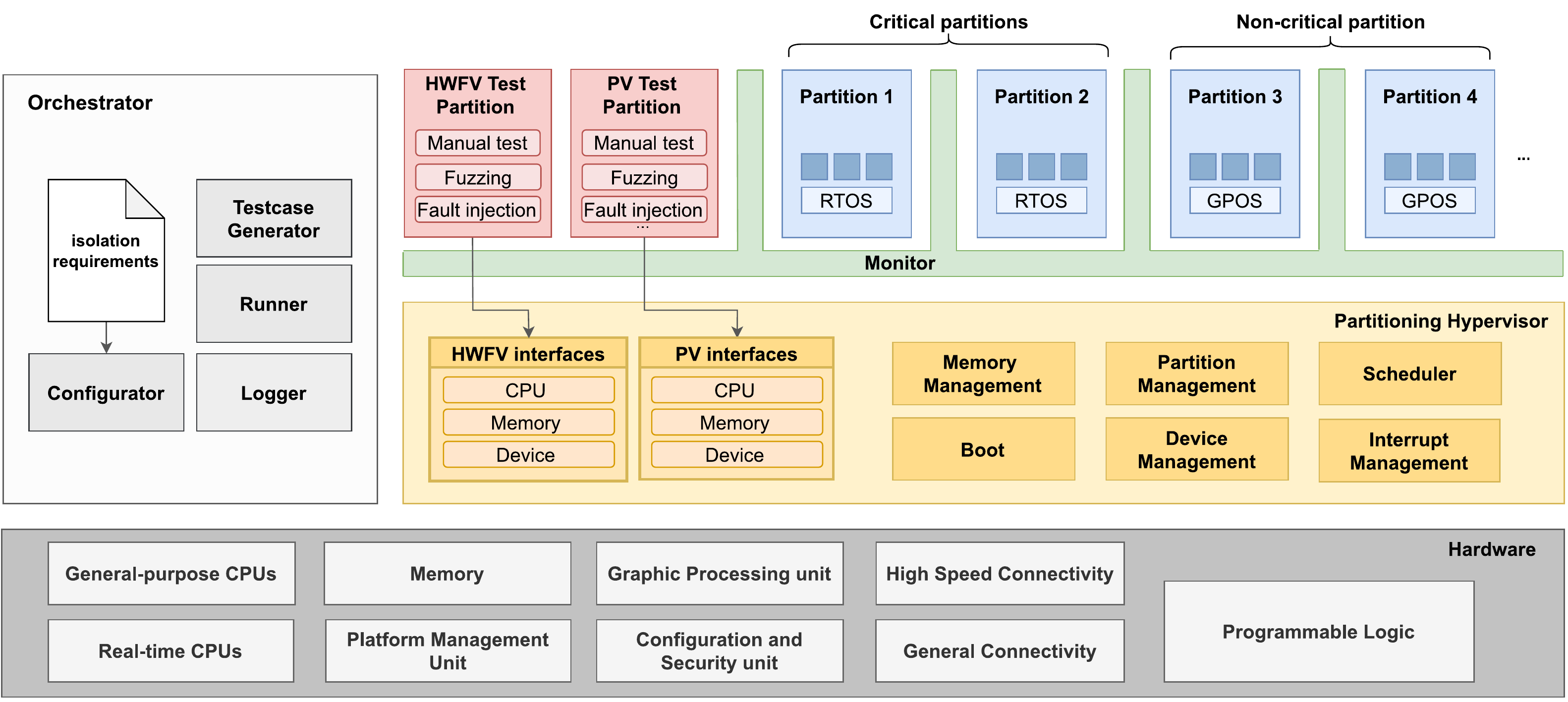}
    \caption{Testing framework design}
    \label{fig:testing_framework_design}
\end{figure*}

Figure ~\ref{fig:testing_framework_design} shows the proposed design for the framework implementation. The architecture components are discussed in detail in the following.

\subsection{Test and regular partitions}
The target hypervisor mounts two types of partitions: \textit{test partitions} and \textit{regular partitions}. \textbf{Test partitions} can be para-virtualized (PV Test Partition) or hardware-assisted full-virtualized (HWFV Test Partition) and will stimulate the hypervisor running \textbf{testing workload}. Different testing techniques can be adopted, such as manually written tests, random tests, dynamic symbolic execution-based tests, automatically generated fuzzing-based tests, fault injection testing, etc. We can map all these techniques to the assurance requirements we find in the standards. We can refer to \tableautorefname{}~\ref{tab:mappingSAR} for mapping. \textit{Regular partitions} are used to run \textit{representative workloads} with different levels of criticality according to the representative scenario deployed on top of the target hypervisor. The hypervisor is assumed to handle these partitions properly while serving the testing workload. 

\begin{table}[!b]
 \renewcommand*{\arraystretch}{2}
\caption{ISO 15408: testing techniques and sars mapping}
\label{tab:mappingSAR}
\resizebox{.5\textwidth}{!}{%
\sffamily 
\footnotesize 
\setstretch{0.90}
\begin{tabular}{@{\extracolsep{\fill}}|>{\raggedright}m{.1\textwidth}|>{\raggedright}m{.3\textwidth}|>{\raggedright}m{.1\textwidth}|}
\hline 

\rowcolor{blue!8}
\centering\textbf{Ref.} & 
\centering\textbf{Security Assurance Requirement} & 
\centering\textbf{Testing Techniques}
\tabularnewline

\hline 
\hline 
\textbf{\emph{[ATE\_FUN]}}
& \textbf{Functional Testing}  \\
The objective is to confirm that the functional testing performed by the developer
Functional testing is performed and documented correctly. The test documentation shall consist of test plans, expected test results, and actual test results. This includes instructions for using test tools and suites, a description of the test environment, test conditions, test data parameters, and values.
&  
End-user-based tests, equivalence tests, boundary values analysis
\tabularnewline

\hline 
\hline 
\textbf{\emph{[ATE\_COV]}}
& \textbf{Coverage Analysis} \\ 
The objective is to confirm that all of the externally visible interfaces (TSFIs), described in the functional specification, have been completely tested.
&
Fuzz testing, dynamic symbolic execution
\tabularnewline

\hline 
\hline 
\textbf{\emph{[ATE\_DPT]}}
& \textbf{Depth Testing} \\ 
The objective is to confirm that all TSF subsystems, described in the ToE design, have been tested. The subsystem descriptions of the TSF provide a high-level description of the internal workings of the TSF. Testing at the level of the ToE subsystems assures that the TSF subsystems behave and interact as described in the ToE design and the security architecture description.
&
Fault injection
\tabularnewline

\hline 
\hline 
\textbf{\emph{[AVA\_CCA]}}
&\textbf{Covert Channel analysis} \\ 
Covert channel analysis is directed toward the discovery and analysis of unintended communications channels that can be exploited to violate the intended TSP.
&
Covert Channel Analysis
\tabularnewline

\hline 
\hline 
\textbf{\emph{[AVA\_SOF]}}
&\textbf{Strength of TOE security functions} \\ 
Strength of function analysis addresses TOE security functions that are realized by a probabilistic or permutational mechanism (e.g. a password or hash function). It is performed to determine whether such functions meet or exceed the claim.
&
Penetration testing
\tabularnewline

\hline 
\hline 
\textbf{\emph{[AVA\_VAN]}}
&\textbf{Vulnerability Analysis} \\ 
"Vulnerability analysis consists of the identification of flaws potentially introduced in the different refinement steps of the development. These potential vulnerabilities are evaluated through penetration testing to determine whether, in practice, they could be exploitable to compromise the security of the TOE."
&
Penetration testing, Fuzz testing, Taint analysis
\tabularnewline

\hline 

\end{tabular}
}
\end{table}

\subsection{FVHW and PV Interfaces}
\textit{Perez-Botero et al.} \cite{perez2013characterizing} perform an extensive analysis of reported vulnerabilities (CVEs) for two general-purpose hypervisors, Xen and KVM. Although a partitioning hypervisor has a smaller attack surface than a general-purpose hypervisor, the above analysis helped define the target interfaces. There are two critical vectors to be considered: the hypercall surface in para-virtualized (PV) systems and the hardware-assisted-full-virtualized (HWFV) surface in full-virtualized systems (e.g., \textit{VMexit handlers} in Intel x86 systems). There is also software-assisted full virtualization, which relies on binary translation. However, it is not used much because current hypervisors take advantage of HWFV to provide better performance. Through PV and HWFV interfaces, it is possible to manage all virtualized resources made available by the hypervisor, such as CPU, memory, and I/O devices. Although para-virtualization outperforms full virtualization in terms of performance \cite{nakajima2007hybrid}, partitioning hypervisors take advantage of HWFV to provide more isolation. For example, Bao \cite{martins2020bao} and Jailhouse \cite{ramsauer2017look} hypervisors leverage ARM VHE \cite{armVHE}, Intel VT-x \cite{serverenabling} CPU virtualization extensions to ensure static partitioning. Additional examples of virtualization extensions that can be used in other CPUs include AMD-V \cite{amd-v} and RISC-V H-extension \cite{sa2021first}. In any case, the framework tries to embrace a broad spectrum of virtualization solutions by including those that are not initially designed for partitioning (see Xen) but can be adapted. In those cases, the para-virtualization surface would continue to exist, could be exploited by an attacker, and therefore must be tested. Using the two types of interfaces described, the framework is generalizable to all hypervisor partitioning solutions.

\subsection{Isolation Monitor}
The \textbf{monitor} component allows us to detect at run-time, during the testing phase, the isolation status provided to regular partitions. It checks whether the testing workload affects the operational state, confidentiality, or integrity of regular partitions. The degree of isolation to be provided by the hypervisor may depend on the standard used for certification, as well as the assurance level to be certified. For this reason, the components to be monitored and the properties to be validated 
should be chosen by a configuration. Generally, logging mechanisms are the main source of information to monitor operation behavior \cite{farshchi2015experience}, but include several limitations, since logs are noisy and lack information on changes in resource states \cite{oliner2012advances}. An effective solution is represented by run-time verification strategies \cite{sanchez2019survey}, which perform checks over events in the system (e.g., after service API calls) to assert whether the resources are in a valid state \cite{bartocci2018introduction}. These checks can be specified as monitoring rules, for example, by using temporal logic and synthesized in a run-time monitor \cite{delgado2004taxonomy, chen2007mop, zhou2014runtime, rabiser2017comparison, cotroneo2018run, cotroneo2015moio}, which allows analyzing whether the testing workload affects the operational state, safety, confidentiality, or integrity of regular partitions. The impact of the monitor component, if it remains in the system after validation, is not negligible. Therefore, this component it must be able to be disabled during production.

\subsection{Orchestrator}
The orchestrator coordinates the testing of the target partitioning hypervisor. First, it configures the testbed via an abstract description (\textbf{configurator}), by providing the isolation monitor with information about which properties and components to monitor. This information may vary depending on the product being tested and the standard considered. The orchestrator also deals with generating test cases (\textbf{test cases generator}), as well as initiating their execution on test partitions. Our framework could potentially handle a wide variety of techniques, ranging from automated fuzzing testing to fault injection, to manually written ad hoc tests. Furthermore, the orchestrator handles startup and any reboots of the target hypervisor during the testing (\textbf{runner}), for example, by using fork servers to parallelize the tests. Finally, it is responsible for maintaining test logs, hypervisor status, and regular partition monitoring logs for analytics purposes (\textbf{logger}).

\section{Related Work}
Over the years several previous works have tried to test virtualization solutions, check their robustness, and reliability, or find security bugs. In some cases, these bugs could affect isolation properties. For example, \textbf{fuzzing techniques} have already been applied at the hypervisor interface to find both bugs and vulnerabilities. MultiNyx \cite{fonseca2018multinyx} proposes a technique for automatically modeling the semantics of complex processor instructions by applying dynamic symbolic execution and combining traces between two different layers (VM and VMM context). Hypercube \cite{schumilo2020hyper} implements a black box fuzzer that automatically discovers and tests different hypervisor interfaces, serving a custom OS to drive the workload. Nyx \cite{schumilo2021nyx} proposes a gray-box coverage-guided fuzzer that relies on coverage retrieved through Intel PT and implements a fast VM reload mechanism to increase fuzzing throughput. Hyperfuzzer \cite{ge2021hyperfuzzer} is a hybrid fuzzer that combines random coverage-guided input mutation, based on Intel PT, with precise input generation based on path constraints derived from dynamic symbolic execution. About \textbf{fault injection testing} targeting virtualization technologies, most studies in the literature proposed methods and tools which only address specific issues of cloud computing-related software. For example, well-known solutions in this field include Fate \cite{gunawi2011fate} and its successor PreFail \cite{joshi2011prefail} for testing cloud-oriented software against faults from the environment, by emulating the unavailability of the network, storage, and remote processes at the API level; similarly, Ju et al. \cite{ju2013fault} test the resilience of cloud infrastructures by injecting crashes (e.g., by killing VMs or service processes), network partitions (by disabling communication between two subnets), and network traffic latency and losses. Cerveira et al. \cite{cerveira2020effects} use fault injection to test isolation among hypervisors and partitions, deliberately introducing CPU / memory corruptions and resource leaks. In contrast to the techniques mentioned above, which try to automate the testing process, both academia and industry have often used \textbf{manually written} tests. NCC Group \cite{michaels2019assessing} tested features such as ASLR, \textit{W\string^X} policy, stack canaries, and heap integrity checks for two major unikernels, demonstrating that security assessment can be done by writing ad hoc tests. The main problem is the effort required to write these tests, in addition to the fact that they are tailored to specific products. Finally, several works have explored the applicability of \textbf{formal methods} directed toward proving the isolation properties of partitioning hypervisors. Baumann et al. \cite{baumann2011proving} verify the memory manager in the PikeOS Separation Kernel through the VCC verifier \cite{cohen2009vcc}. Richards \cite{richards2010modeling} demonstrated verification of the fault containment property of the INTEGRITY-178B Separation Kernel used in avionics. Heitmeyer et al. \cite{heitmeyer2006formal} demonstrate the formalization of a data separation property and its verification for an embedded device. To date, the use of formal methods appears to be the most comprehensive way to verify the isolation requirements of these solutions. The major limitations, however, are due to excessive cost in the analysis and non-applicability for solutions larger than 10/15K LOC. In addition, although they are mandated techniques for certification of high assurance levels (e.g. EAL6-EAL7 in CC), they are not required for intermediate levels. Beyond formal methods, the other techniques mentioned are not sufficient when used alone. Although many of them appear to be applicable to partitioning hypervisors, (1) none of them specifically consider certification requirements related to isolation, (2) individually they do not exhaustively cover all partitioning hypervisor components, and (3) they are not generalizable to different partitioning hypervisor solutions. For these reasons, it is instead appropriate to use these techniques in combination within the proposed framework.

\section{Conclusion}

This work provides a methodological framework to evaluate the isolation level of a partitioning hypervisor solution. The framework is generalizable to different products and explicitly considers the isolation requirements of security- and safety-related standards. By analyzing the isolation requirements provided in different standards, the hypervisor properties to be validated during the certification process were well-defined. In addition, the SKPP analysis provided an example mapping between functional mechanisms and certification functional requirements, as well as a mapping between existing testing techniques and certification assurance requirements. These mappings must be considered during the use of the framework to generate evidence about the coverage of a certain set of certification requirements. An industrial player can use our framework to assess the isolation properties of different existing products and choose the one that best suits their needs. On the other hand, a developer might be interested in verifying his own product to place it in a certifiable context and provide a certification package.

\section*{Acknowledgments}
This work has been supported by the project COSMIC of UNINA DIETI.

\bibliographystyle{IEEEtran}
\bibliography{main.bib}

\end{document}